\documentclass{epl}
\usepackage{graphicx} 
\usepackage{psfrag}
\usepackage{fancyhdr}  
\usepackage{bm}

\title{Electromagnetic instability of the Thomson Problem}
\shorttitle{The Thompson Problem}        

\author{Jayme De Luca\thanks{E-mail: \email{deluca@df.ufscar.br}},
Savio B. Rodrigues\thanks{E-mail: \email{savio@dm.ufscar.br}}
\and Yan Levin\thanks{E-mail: \email{levin@if.ufrgs.br}}}
\institute{$^*$Departamento de F\'{\i}sica,
Universidade Federal de S\~ao Carlos, Rod. Washington Luiz km 235,
13565-905,  Caixa Postal 676, S\~ao Carlos, SP, Brazil\\
$^{**}$Departamento de Matem\'atica,
Universidade Federal de S\~ao Carlos, Rod. Washington Luiz km 235,
13565-905,  Caixa Postal 676, S\~ao Carlos, SP, Brazil\\
$^{***}$Instituto de F\'{\i}sica,
  Universidade Federal do Rio Grande do Sul \\
  Caixa Postal 15051, CEP 91501-970, Porto Alegre, RS, Brazil}

\pacs{64.60.Cn}{Order-disorder transformations; statistical mechanics of
model systems}
\pacs{71.10.-w}{Electron gas-theories and models}

\date{\today}
 
\begin{document}    
\maketitle

\begin{abstract}
The classical   
Thomson problem of  $n$ charged particles 
confined to the surface of a sphere of radius $a$ is analyzed within 
the Darwin approximation of electrodynamics. For 
$n<n_c(a)$ the ground state corresponds to a hexagonal 
Wigner crystal with
a number of topological defects.  However, if $n>n_c(a)$ the Wigner 
lattice is unstable with respect to small perturbations
and the ground state becomes spontaneously magnetized for finite $n$.
\end{abstract}

The Thomson problem, finding the ground state of electrons inside a sphere
with a uniform neutralizing background,  
has a time honored position in the history of 
modern physics~\cite{SaKu97,CaRiSt98,MaDa93,BeSh94,PeMo99,MeHoKr01,BoCaNe02}.
The original question was posed by J.J. Thomson~\cite{Th04} after his 
discovery of the electron in $1897$.   
Thomson  conjectured that the knowledge of the positions of the
electrons inside the atoms is essential  to understanding the
regularity of the chemical elements in the periodic table.  
At the time, however, proton still had to wait $14$ years to be discovered, 
so in order to keep his atom neutral, Thomson was forced to introduce a 
uniform neutralizing background.
The model became known as the ``plum pudding'' atom and the question
that needed to be answered was:  
What are the positions of the electrons inside a uniformly (positively) 
charged sphere?  
Surprisingly, after more than a century this problem
still has no general solution.

If the background charge is made to vanish,
the electrostatic energy 
will be a minimum only if all the electrons are located 
at the surface. This is a general consequence of the   
Earnshaw theorem~\cite{Ea42} which precludes existence of a 
stable equilibrium
with purely electrostatic interactions.
Curiously, the Coulomb potential is precisely on the border line where this
behavior is possible.  If instead of $1/r$, the electrons would
interact by a $1/r^{1+\epsilon}$ potential 
with $\epsilon >0$, the bulk occupation
of the sphere would be energetically favorable 
for a sufficiently large number of electrons~\cite{LeAr03}.  
Unfortunately, even the restricted surface 
Thomson problems remains unsolved for
an arbitrary number of electrons~\cite{Al97,PeDo97}.

In this letter we will show that if the 
relativistic  corrections to the Coulomb law are properly taken
into account, even our intuitive picture of the ground state
as consisting of stationary particles located at fixed positions on
the surface of a sphere must be abandoned. Instead, we find
that for sufficiently large electron density, the energy is minimized by
the particles undergoing a coherent motion and  
the sphere becomes spontaneously magnetized!

The starting point for our analysis is the well known Darwin 
Lagrangian~\cite{Da20,LaLi62,TrKo68,KrHa62,Kr69,Kr74,Es96,Es97,ApAl98,MeLu00,MeLu01},
which takes into account the relativistic corrections to the Coulomb law
resulting from the particle motion,
\begin{equation}
\label{1}
L=- m c^2\sum_i \sqrt{1-\frac{v_i^2}{c^2}}-
\frac{1}{2}\sum_{i \ne j} \frac{q_i q_j}{r_{ij}}+
\frac{1}{4 c^2}\sum_{i \ne j} \frac{q_i q_j}{r_{ij}}\left[ \bm v_i\cdot \bm v_j+(\bm v_i \cdot \hat{\bm r}_{ij})(\bm v_j\cdot \hat{\bm r}_{ij} )\right]\;.
\end{equation}
Eq.~(1) is correct to order $v^2/c^2$.
The velocity-dependent correction to the Coulomb energy arises 
from the electromagnetic coupling between
the moving particles.  Since the Lagrangian (\ref{1}) does not contain  
explicit time dependence, the energy of the system  
\begin{equation}
\label{2}
E=\sum_i \bm v_i \cdot \frac{\partial L}{\partial \bm v_i} -L \;
\end{equation}
is a constant of motion,
\begin{equation}
\label{4}
E=\sum_i \frac{m c^2 }{\sqrt{1-\frac{v_i^2}{c^2}}}+
\frac{1}{2}\sum_{i \ne j} \frac{q_i q_j}{r_{ij}}+
\frac{1}{4 c^2}\sum_{i \ne j} \frac{q_i q_j}{r_{ij}}\left[ \bm v_i\cdot \bm v_j+(\bm v_i \cdot  \hat{\bm r}_{ij})(\bm v_j\cdot  \hat{\bm r}_{ij} )\right]\;.
\end{equation}
The ground state for $n$ electrons on the surface of a sphere of radius
$a$ is then determined by the minimization of Eq. (\ref{4}).

We note that if the terms of order $1/c^2$ are neglected, we recover the
classical formulation of the Thomson problem in which the electromagnetic
coupling between the electrons is purely of the Coulomb form. In this
case, the velocity dependent contribution to the Hamiltonian is 
positive or zero, and the ground state corresponds to
stationary particles residing at fixed positions on the surface of the
sphere.  For large $n$, this structure resembles a 
hexagonal Wigner crystal containing some topological defects.  
In general, however, the $1/c^2$ terms can not be omitted and a full
minimization of Eq. (\ref{4}) must be performed.  
To proceed, it is convenient
to rewrite the energy in adimensional form. Defining the reduced
displacement and velocity as $r^*=r/a$ and $v^*=v/c$, the reduced
energy becomes
\begin{equation}
\label{5}
E^* \equiv 
E \frac{r_e}{q^2}=\sum_i\frac{1}{\sqrt{1- v_i^{*2}}}+
\frac{1}{2 a^*}\sum_{i \ne j} \frac{1}{r^*_{ij}}+
\frac{1}{4 a^*}\sum_{i \ne j} \frac{1}{r^*_{ij}}\left[ \bm v^*_i\cdot \bm v^*_j+(\bm v^*_i \cdot  \hat{\bm r}_{ij})(\bm v^*_j\cdot  \hat{\bm r}_{ij} )\right] \;,
\end{equation}
where $r_e \equiv q^2/m c^2$ is the classical electron radius and
$a^*=a/r_e$. 

It is convenient to work in the spherical coordinate system with unit vectors 
$\hat{\bm n},\hat{\bm\theta},\hat{\bm\phi}$.
The reduced velocity of electron $i$ on the surface of the
sphere is then $\hat{\bm v}^*_i= \hat{\bm \theta} v_{\theta i}+\hat{\bm \phi} v_{\phi i}$.

\begin{figure}[h]
\begin{center}
\psfrag{PHI}{\large $ \phi$}
\psfrag{THETA}{\large $ \theta$}
\includegraphics[width=10cm]{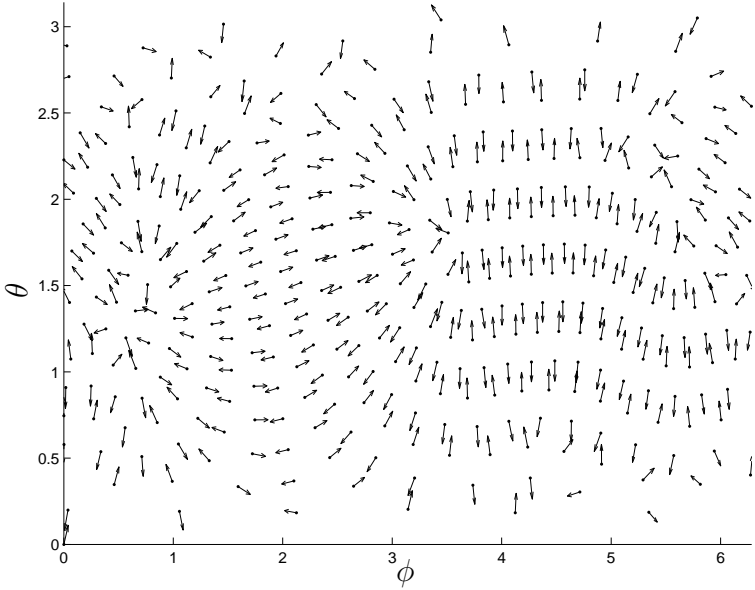}
\end{center}
\caption{The velocity field of electrons in the supercritical region,
$n=400$ and $a = 0.7 a_c$.}
\label{fig4}
\end{figure}

Minimization of  $E^*$, Eq.~(\ref{5}), is 
performed using a general purpose 
quasi-Newton method where the Hessian update is given by the 
BFGS formula~\cite{Lu84}. Gradients are computed analytically. A line 
search with cubic fit is used with the additional safeguard 
against evaluations beyond light speed. The procedure is 
highly non-trivial.  In fact it is known that already
for the classical Thomson problem, in the 
absence of relativistic corrections,
there exists an exponentially large number of 
metastable states~\cite{ErHo95}.  
Thus, it is
quite unlikely that any  minimization procedure will be able to
locate the exact ground state for a large number of electrons.  This however,
is not of great importance since the metastable states have energies
very close to that of the exact ground state~\cite{ErHo95,LeAr03}. 
Performing the minimization of $E^*$ we find that for reduced surface charge
density $\sigma^*=n/a^{*2}$ such 
that  $\sigma^*<\sigma_c^*$ (subcritical region), 
the electrons form a stationary
Wigner crystal with some topological defects.  Above the critical charge
density $\sigma^*>\sigma_c^*$ (supercritical region), 
the Wigner crystal, however, becomes unstable
and a new ground state with moving electrons is formed.  In Fig. \ref{fig4}
we show the characteristic distribution of particles in this new ground
state.  The arrows indicate the relative magnitude and direction
of the particle velocities.  The figure 
shows bands of correlated antiferromagnetic velocities  
that try to adapt to the topology of the sphere.
We stress that
when the instability occurs, $v_i/c \ll 1$ for all the particles,
so that the Darwin Lagrangian remains valid, up to quantum corrections. 
The melting of the Wigner crystal is an example of a classical 
zero temperature phase transition. 

To better understand the nature of the  instability of the 
Wigner lattice, it is convenient to rewrite the Darwin energy
in a matrix form. Defining a $2 n$ component velocity vector
$\bm V= \{{\bm v^*_1},{\bm v^*_2},... {\bm v^*_n}\} $, 
Eq.~(\ref{5}) can be rewritten to order $1/c^2$ as
\begin{equation}
\label{14}
E^*  \approx  n+\frac{1}{2}{\bm V^T}\, {\rm \bf I}\, {\bm V}+
\frac{1}{2 a^*}{\bm V^T}\, {\rm \bf D}\, {\bm V} +
\frac{3}{8}\sum v_i^{*4}+
\frac{1}{2 a^*}\sum_{i \ne j} \frac{1}{r^*_{ij}}\;,
\end{equation}
where ${\rm \bf  I}$ is a $2n \times 2n $ identity 
matrix and ${\rm \bf D}$ is a position dependent 
matrix constructed  from the last term 
of Eq.~(\ref{5}).  The quadratic term in velocity is non-negative
if all the eigenvalues of the matrix
\begin{equation}
\label{15}
{\rm \bf  A}={\rm \bf  I}+\frac{1}{a^*}{\rm \bf D} \;,
\end{equation}
are positive.  In this case the ground state will have $\bm V=0$ and the
electrons will be organized into a Wigner crystal.  On the other hand,
as soon as one of the eigenvalues of ${\rm \bf  A}$ becomes 
negative, the Wigner lattice will lose stability, and a new ground state,
with energy below that of the Wigner crystal will be established.  
The phase transition  occurs when $\lambda_{min}^{\rm \bf  A}=0$, where 
$\lambda_{min}^{\rm \bf  A}$ is the minimum 
eigenvalue of the matrix ${\rm \bf  A}$.

It is important to note that the energetic bifurcation 
of Eq. (\ref{14}) is simultaneous with the 
dynamical instability of the Wigner lattice.  
If the Euler-Lagrange equations of motion
are linearized around the stationary positions
of the Wigner lattice, one can show that
the Lyapunov instability occurs precisely when 
${\rm \bf  A}$ loses convexity.
Unfortunately, 
in the supercritical region,
the equations of motion 
are differential-algebraic and due to  the singularity 
of  ${\rm \bf  A}$ are very difficult to 
integrate numerically~\cite{BrCaPe89}.

To determine the critical charge concentration at which the Wigner
crystal loses stability, we adopt the following procedure.  
For a given number of electrons $n$, the 
Coulomb energy is minimized to determine the positions of all the particles.
For purely Coulombic interactions, the ground state location of the
electrons is independent of the size of the sphere, since $a$ scales  
out of the expression for the electrostatic energy. Once the ground
state coordinates are known, the eigenvalues $\lambda^{\rm \bf  D}$
of the matrix ${\rm \bf  D}$ can be calculated numerically.
The criticality condition  $\lambda_{min}^{\rm \bf  A}=0$ is then
equivalent to the requirements that $\lambda_{min}^{\rm \bf  D}=-a^*$ .
In Fig. \ref{2} we show the result of this procedure.
\begin{figure}[h]
\begin{center}
\psfrag{ac}{\huge $ a_c$}
\psfrag{n}{\huge $ n$}
\includegraphics[width=8cm,angle=270]{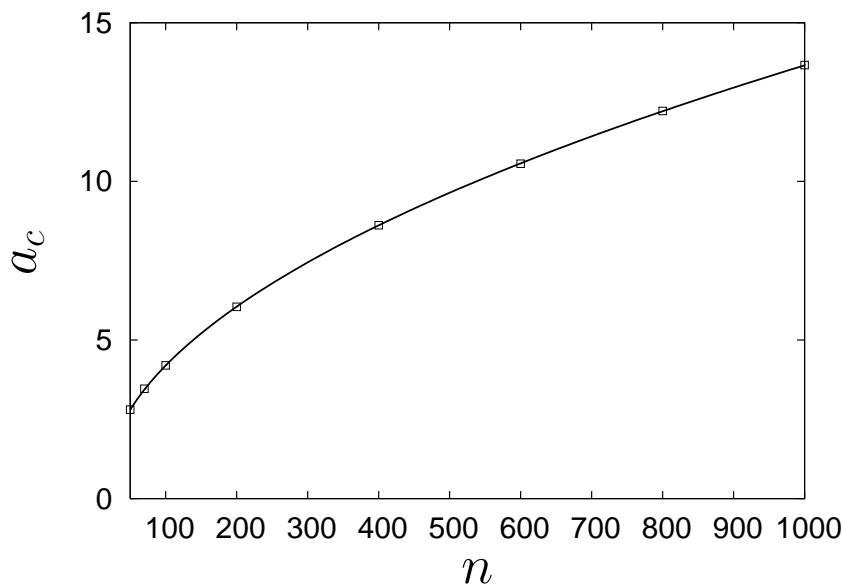}
\end{center}
\caption{The critical radius as a function of $n$.  The solid line
is a fit given by Eq.~(\ref{16a})}
\label{fig2}
\end{figure}

The points in Fig. \ref{fig2} can be very well fitted by
\begin{equation}
\label{16a}
a_c^*= 0.4323\, n^{\frac{1}{2}}-\frac{12.680}{n}\;.
\end{equation}
Eq.~({\ref{16a}}) implies existence of a well defined thermodynamic 
limit for the phase transition, $\lim n \rightarrow \infty$, 
$a \rightarrow \infty$ and  
$\sigma^*_c \rightarrow 5.35$.

\begin{figure}[h]
\begin{center}
\psfrag{mu}{\large $ \mu^* n^{2/5}$}
\psfrag{ra}{\large $\sigma^*-\sigma_c^*$}
\includegraphics[width=8cm, angle=270]{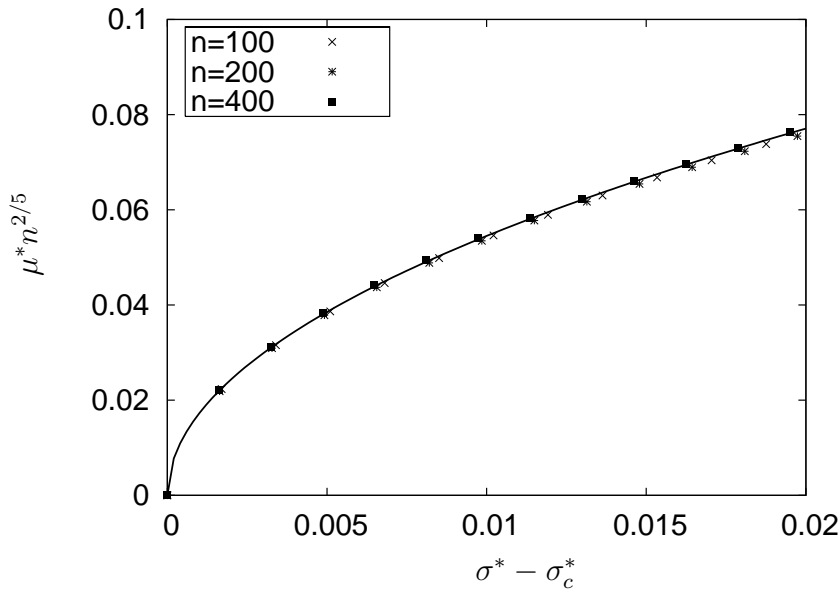}
\end{center}
\caption{Magnetic moment in the supercritical region, the data points
are the averages over five random initial configurations.}
\label{fig3}
\end{figure}

To further explore the nature of the ground state for $\sigma>\sigma_c$
we define an order parameter
\begin{equation}
\label{16}
\bm \mu^*= \sum_i r_i^* \times \bm v_i^* \;.
\end{equation}
Clearly, $\bm \mu^*$  is just proportional to the total magnetic moment
of the sphere.  In the subcritical region, the electron
velocities are zero and $\bm \mu^*=0$. The value of the
magnetic moment in the supercritical region
is plotted in 
Fig. \ref{fig3}.  We find that if the magnetic moment is scaled with
$n^{-2/5}$ and is plotted as a function of the reduced surface charge
concentration $\sigma^*-\sigma_c^*$, all the points  for different values of
$a$ and $n$ fall on the
same universal curve,
\begin{equation}
\label{17}
g(x)=0.545\, x^{1/2}\;.
\end{equation}
Thus, although locally the orientation of the velocity vectors is
antiferromagnetic, globally the symmetry is broken and the
sphere acquires a net magnetic moment. 
The magnetic moment is sub-extensive
and vanishes with a square root singularity as  
$\sigma \rightarrow \sigma_c^+$.

\begin{figure}[h]
\begin{center}
\psfrag{e}{\large $ \Delta E^*/n$}
\psfrag{n}{\large $\sigma^*$}
\includegraphics[width=8cm,angle=270]{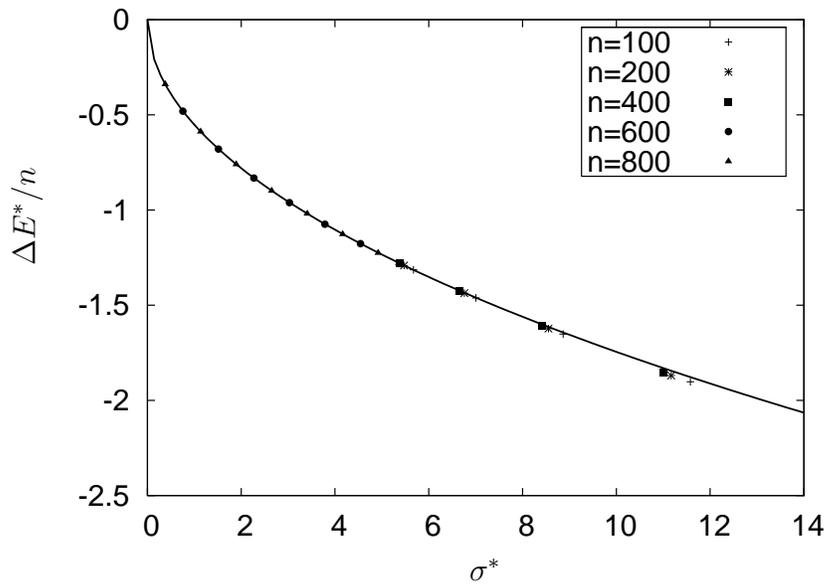}
\end{center}
\caption{The $\Delta E^*/n$ as a function of $\sigma^*$ for various
combinations of $n$ and $a$. }
\label{fig1}
\end{figure}

We next proceed to study the extensive property of 
the electromagnetic energy $E^*$.  For $\sigma < \sigma_c$
a very accurate expression
for the ground state  of the
Thomson problem~\cite{ErHo91,Le02}
can be obtained using a simple argument.
Consider a uniformly {\it positively} charged spherical shell on which
move $n$ electrons. This problem defines a spherical one-component
plasma (SOCP). The electrostatic (Coulomb) energy can be written as  
\begin{equation}
\label{8} 
F_{SOCP}=E_C+
\frac{q^2 n^2}{2 a} -
\frac{q^2  n^2}{a}  \;.
\end{equation}
The first term is the Coulomb  energy of interaction between $n$ electrons
on the surface of the sphere,
the second term is the self energy of the  positive background charge,
and the third term is the energy of interaction between $n$ electrons
and the background.  At zero temperature, the classical SOCP will freeze
into a hexagonal Wigner crystal (with some topological defects)  
whose energy is
\begin{equation}
\label{9} 
F_{SOCP}=-M \frac{q^2 n}{d}\;,
\end{equation}
where $M$ is the Madelung constant and $d$ is the characteristic size of
the Wigner-Seitz cell, $\pi d^2 n =4 \pi a^2$.  Combining Eqs. (\ref{8}) and
(\ref{9}) we arrive at a very simple expression for the Coulomb energy
of $n$ electrons on the surface of the sphere 
\begin{equation}
\label{10}
E_C= \frac{n^2 q^2}{2  a}-M \frac{q^2 n^{3/2} }{ 2 a }\;.
\end{equation}
Eq.~(\ref{10}) with $M=1.1046$ 
gives a very accurate fit to the ground state energy of
the surface Thomson problem with purely 
Coulomb interactions~\cite{ErHo91,LeAr03}. 
Note that for a 
planar OCP~\cite{GaChCh79} $M=1.1061$, so that the topological
defects affect very little the value of the Madelung constant.
It is also important
to notice that although $E_C$ is not extensive,
\begin{equation}
\label{11}
\Delta E_C \equiv \frac{1}{q^2}\left(E_C - \frac{n^2 q^2}{2 a}\right) \;
\end{equation}
is. Therefore, if  $\Delta E_C/n$ is plotted  as a function of $n/a^2$ 
for different combinations of $n$ and $a$, 
all points should fall onto one universal curve,
\begin{equation}
\label{12}
f(x)=-\frac{M }{2 }\sqrt x\;.
\end{equation}
We can now check if this universality also holds for the
Thomson problem with the Darwin coupling between the particles.
That is if  
\begin{equation}
\label{13}
\Delta E^* \equiv E^* - 1 - \frac{n^2}{2 a^*}  \;,
\end{equation}
is such that 
$\Delta E^*=n f(\sigma^*)$, with $f(x)$ given by Eq.~(\ref{12}).  
In Fig.\ref{fig1} $\Delta E^*/n$  is plotted as a function
of $\sigma^*$ for various combinations of $n$ and $a^*$.  
It is quite surprising that even in the supercritical region
$\sigma^*>5.4$, the deviation of $E^*$ 
from the energy of a stationary Wigner 
crystal remains very small.  This is in spite of the
fact that for $\sigma>\sigma_c$, the velocities
of individual particles can be quite large. Evidently, the
local antiferromagnetic ordering of the velocity vectors leads
to significant cancellations which diminish the overall contribution
of the Darwin term to the total energy.

We have shown that if the relativistic corrections are
taken into account, 
the classical Thomson problem of the electrons confined
to the surface of a sphere exhibits an electromagnetic instability. 
While for $\sigma<\sigma_c$, the ground state of electrons 
is a Wigner crystal with some topological defects, 
for $\sigma>\sigma_c$, 
the Wigner lattice is unstable and a 
small perturbation can make the system evolve to a new ground state.  This
ground state is characterized by a local 
antiferromagnetic order~\cite{ApAl98,MeLu00,MeLu01}, 
but finite net magnetic moment.  The surface charge 
concentration at the phase transition
has a well defined thermodynamic limit $n \rightarrow \infty$,
$a \rightarrow \infty$, while $\sigma^*_c \rightarrow 5.4$. 
This surface charge density, however, is so large that quantum effects
must be taken into account~\cite{AlAp00,ApAl99}.  The relativistic
corrections to the Coulomb energy should not, therefore, affect the
stability of a normal plasma. 

This work was supported in part by the Brazilian agencies
CNPq and FAPESP.


\begin{thebibliography}{10}

\bibitem{SaKu97}
E.~B. Saff and A.~B.~J. Kuijlaars, Math. Inteligencer {\bf 19},  5  (1997).

\bibitem{CaRiSt98}
L. Candido, J.~P. Rino, N. Studart, and F.~M. Peeters, J. Phys.: Condens.
  Matter {\bf 10},  11627  (1998).

\bibitem{MaDa93}
C.~J. Marzec and L.~A. Dat, Biophys. J. {\bf 65},  2559  (1993).

\bibitem{BeSh94}
B. Berger, P.~W. Shor, L. Tuckerkellogg, and J. King, Proc. Natl. Acad. Sci.
  U.S.A. {\bf 91},  7732  (1994).

\bibitem{PeMo99}
A. Perez-Garrido and M.~A. Moore, Phys. Rev. B {\bf 60},  15628  (1999).

\bibitem{MeHoKr01}
R. Messina, C. Holm, and K. Kremer, Phys. Rev. E {\bf 64},  021405  (2001).

\bibitem{BoCaNe02}
M. Bowick, A. Cacciuto, D.~R. Nelson, and A. Travesset, Phys. Rev. Lett. {\bf
  89},  185502  (2002).

\bibitem{Th04}
J.~J. Thomson, Philos. Mag. {\bf 7},  237  (1904).

\bibitem{Ea42}
S. Earnshaw, Trans. Camb. Phil. Soc. {\bf 7},  97  (1842).

\bibitem{LeAr03}
Y. Levin and J.~J. Arenzon, Europhys. Lett. {\bf 63},  415  (2003).

\bibitem{Al97}
E.~L. Altschuler~et al., Phys. Rev. Lett. {\bf 78},  2681  (1997).

\bibitem{PeDo97}
A. Perez-Garrido, M.~J.~W. Dodgson, and M.~A. Moore, Phys. Rev. Lett. {\bf 79},
   1417  (1997).

\bibitem{Da20}
C.~G. Darwin, Philos. Mag. {\bf 39},  537  (1920).

\bibitem{LaLi62}
L.~D. Landau and E.~M. Lifshitz, {\em The theory of classical fields} (Pergamon
  Press, Oxford, 1962).

\bibitem{TrKo68}
B.~A. Trubnikov and V.~V. Kosachev, Sov. Phys. JETP {\bf 27},  501  (1968).

\bibitem{KrHa62}
J.~E. Krizan, Phys. Rev. {\bf 128},  2916  (1962).

\bibitem{Kr69}
J.~E. Krizan, Phys. Rev. {\bf 177},  376  (1969).

\bibitem{Kr74}
J.~E. Krizan, Phys. Rev. A {\bf 10},  298  .

\bibitem{Es96}
H. Essen, Phys. Rev. E {\bf 53},  5228  (1996).

\bibitem{Es97}
H. Essen, Phys. Rev. E {\bf 56},  5858  (1997).

\bibitem{ApAl98}
W. Appel and A. Alastuey, Physica A {\bf 252},  238  (1998).

\bibitem{MeLu00}
V. Mehra and J. De~Luca, Phys Rev. E. {\bf 61},  1199  (2000).

\bibitem{MeLu01}
V. Mehra and J. De~Luca, Phys. Rev. B {\bf 64},  085306  (2001).

\bibitem{Lu84}
D.~G. Luenberger, {\em Linear and {N}onlinear {P}rogramming} (Addison-Wesley
  Publishing Company, Reading, Massachusetts, 1984).

\bibitem{ErHo95}
T. Erber and G.~M. Hockney, Phys. Rev. Lett. {\bf 74},  1482  (1995).

\bibitem{BrCaPe89}
K. Brenan, S.~L. Campbell, and L.~R. Petzold, {\em Numerical {S}olution of
  {I}nitial-{V}alue {P}roblems in {D}ifferential-{A}lgebraic {E}quations}
  (Elsevier, New York, 1989).

\bibitem{ErHo91}
T. Erber and G.~M. Hockney, J. Phys. A {\bf 24},  L1369  (1991).

\bibitem{Le02}
Y. Levin, Rep. Prog. Phys. {\bf 65},  1577  (2002).

\bibitem{GaChCh79}
R.~C. Gann, S. Chakravarty, and G.~V. Chester, Phys. Rev. B {\bf 20},  326
  (1979).

\bibitem{AlAp00}
A. Alastuey and W. Appel, Physica A {\bf 276},  508  (2000).

\bibitem{ApAl99}
W. Appel and A. Alastuey, Phys. Rev. E {\bf 59},  4542  (1999).

\end{thebibliography}

\end{document}